\documentclass[lettersize,journal]{IEEEtran}
\usepackage[utf8x]{inputenc} 
\usepackage{amsmath,amsfonts}
\usepackage{algorithmic}
\usepackage{algorithm}
\usepackage{array}
\usepackage{textcomp}
\usepackage{stfloats}
\usepackage{url}
\usepackage{verbatim}
\usepackage{graphicx}
\usepackage{cite}
\usepackage{graphicx}
\usepackage{amsmath}
\usepackage{amssymb}
\usepackage{booktabs} 
\usepackage{subcaption}
\usepackage{graphicx}
\usepackage{float}
\usepackage{bm}
\usepackage{times}
\usepackage{caption}

\usepackage{makecell}
\usepackage{multicol}
\usepackage{multirow}
\usepackage{color}
\usepackage{booktabs}
\usepackage{float}
\usepackage{amssymb}
\usepackage{wasysym}
\usepackage{ulem}
\usepackage{pifont}
\usepackage[thicklines]{cancel}
\usepackage[pagebackref,breaklinks,colorlinks]{hyperref}
\usepackage[capitalize]{cleveref}
\crefname{section}{Sec.}{Secs.}
\Crefname{section}{Section}{Sections}
\Crefname{table}{Table}{Tables}
\crefname{table}{Tab.}{Tabs.}

\begin{document}

\title{ShiftLIC: Lightweight Learned Image Compression with Spatial-Channel Shift Operations}
\author{Youneng Bao, Wen Tan, Chuanmin Jia, Mu Li$^{*}$, \\ Yongsheng Liang$^{*}$,~\IEEEmembership{Member,~IEEE} and Yonghong Tian,~\IEEEmembership{Fellow,~IEEE,}
\thanks{Y. Bao, W. Tan and Y. Liang are with the School of Electronics and Information Engineering, M. Li are with the School of Computer Science, 
Harbin Institute of Technology, Shenzhen, Guangdong, China (e-mail: ynbao@stu.hit.edu.cn; tanwen150548@163.com;limuhit@gmail.com;)}
\thanks{Y. Liang is also with the College of Big Data and Internet, Shenzhen Technology University, Shenzhen, 518118, Guangdong, China}
\thanks{Chuanmin Jia is with the Wangxuan Institute of Computer Technlogy (WICT), Peking University, Beijing 100871, China (e-mail: cmjia@pku.edu.cn).}
\thanks{Y. Tian is with the Peng Cheng Laboratory, Shenzhen, China (e-mail:
tianyh@pcl.ac.cn).}

}

\markboth{Journal of \LaTeX\ Class Files,~Vol.~14, No.~8, August~2022}%
{Shell \MakeLowercase{\textit{et al.}}: A Sample Article Using IEEEtran.cls for IEEE Journals}

\maketitle

\begin{abstract}
    Learned Image Compression (LIC) has attracted considerable attention due to their outstanding rate-distortion (R-D) performance and flexibility. However, the substantial computational cost poses challenges for practical deployment. The issue of feature redundancy in LIC is rarely addressed. Our findings indicate that many features within the LIC backbone network exhibit similarities. 
    This paper introduces ShiftLIC, a novel and efficient LIC framework that employs parameter-free shift operations to replace large-kernel convolutions, significantly reducing the model's computational burden and parameter count. Specifically, we propose the Spatial Shift Block (SSB), which combines shift operations with small-kernel convolutions to replace large-kernel. This approach maintains feature extraction efficiency while reducing both computational complexity and model size. To further enhance the representation capability in the channel dimension, we propose a channel attention module based on recursive feature fusion. This module enhances feature interaction while minimizing computational overhead. Additionally, we introduce an improved entropy model integrated with the SSB module, making the entropy estimation process more lightweight and thereby comprehensively reducing computational costs.
    Experimental results demonstrate that ShiftLIC outperforms leading compression methods, such as VVC Intra and GMM, in terms of computational cost, parameter count, and decoding latency. Additionally, ShiftLIC sets a new SOTA benchmark with a BD-rate gain per MACs/pixel of -102.6\%, showcasing its potential for practical deployment in resource-constrained environments.  The code is released at \url{https://github.com/baoyu2020/ShiftLIC}.

\end{abstract}


\section{Introduction}
\label{sec:intro}
\IEEEPARstart{I}{mage} compression remains a cornerstone issue in multimedia transmission and storage. The community has increasingly focused on lossy image compression using deep neural networks, owing to its superior rate-distortion (R-D) performance.
Currently, neural network-based image compression algorithms have surpassed traditional handcrafted standards in performance, such as JEPG\cite{JPEG}, JPEG2000\cite{marcellin2000overview}, HEVC Intra\cite{sullivan2012overview}, VVC\cite{bross2021overview} and AVS\cite{AVS3TMM,maEvolutionAVSVideo2022} as referenced in \cite{iWave2020TMM, bao2023nonlinear, he2022elic,mazhan2023wacv,guo2023efficient}.
However, these state-of-the-art (SOTA) Learned Image Compression (LIC) algorithms are computationally intensive and parameter-heavy, posing challenges to practical deployment. Therefore, developing lightweight learned image compression algorithms is crucial.\par

\begin{figure}[t]
    \centering 
    \includegraphics[width=9cm]{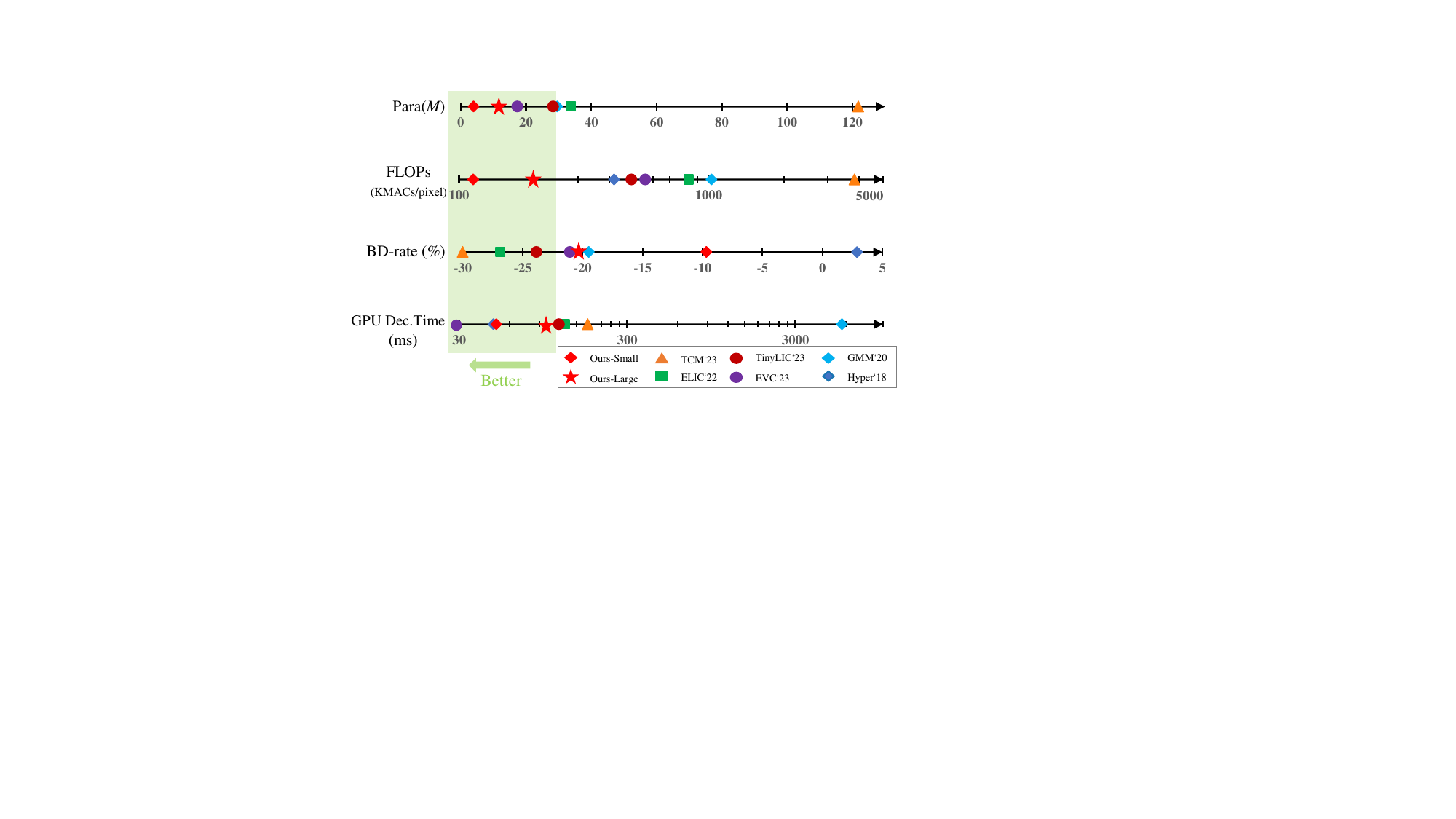}
    \caption{\textbf{A comprehensive comparison of LIC performance across four dimensions.} 
    R-D performance is quantified using BD-rate, which is anchored by BPG. Computational complexity is measured in terms of KMACs (thousand multiply-accumulate operations) per pixel. Parameter complexity is measured as the total number of parameters of the model. Our models achieve an excellent trade-off between performance and complexity. 
    Representative work, such as Hyper’18\cite{balle2018variational}, TCM’23\cite{liu2023learned}, GMM’20\cite{cheng2020learned}, ELIC’22\cite{he2022elic},EVC’23\cite{guo-hua2023evc} and TinyLIC’23\cite{lu2022high}, is compared.
    Refer to Table \ref{table:quantitative} for the results of additional methods. } 
    \label{fig:rd-c}
    \vspace{-0.5cm}
    \end{figure}

Regarding the acceleration of LIC, prior efforts have mainly focused on two areas: the development of efficient and rapid parallel decoding algorithms, and the weight quantization or pruning of LIC models.
Parallel decoding algorithms\cite{He_2021_CVPR} significantly reduce the decoding latency introduced by autoregressive algorithms\cite{minnen2018joint,chengtong2021tip}, but some channel-wise entropy models\cite{Channel-wise, lu2022high, liu2023learned} add additional parameters and computational load. Weight quantization\cite{Shi2023Rate, le2022mobilecodec, van2023mobilenvc} and model pruning\cite{guo-hua2023evc} both lead to a reduction in parameter count and computational acceleration. Weight quantization and model pruning reduce parameter counts and accelerate computations. However, these techniques lead to performance degradation in image compression tasks and generally necessitate specialized hardware for acceleration.
 
Lightweight algorithm design involves building efficient models from the ground up, providing greater flexibility across various platforms and can be used in conjunction with the aforementioned acceleration strategies.\par

The key strategies for lightweight algorithm design targeting convolutional neural networks include increasing group convolutions or reducing the size of the  kernels. 
Notable research in this domain include SqueezeNet\cite{iandola2016squeezenet}, MobileNet\cite{howard2017mobilenets}, ShuffleNet\cite{zhang2018shufflenet}, EfficientNet\cite{tan2019efficientnet}, among others. These algorithms are designed for high-level semantic tasks. However, the development of lightweight algorithms specifically for image compression has not yet been thoroughly explored. \par
\begin{figure*}[t]
    \centering 
    \includegraphics[width=1.0\textwidth]{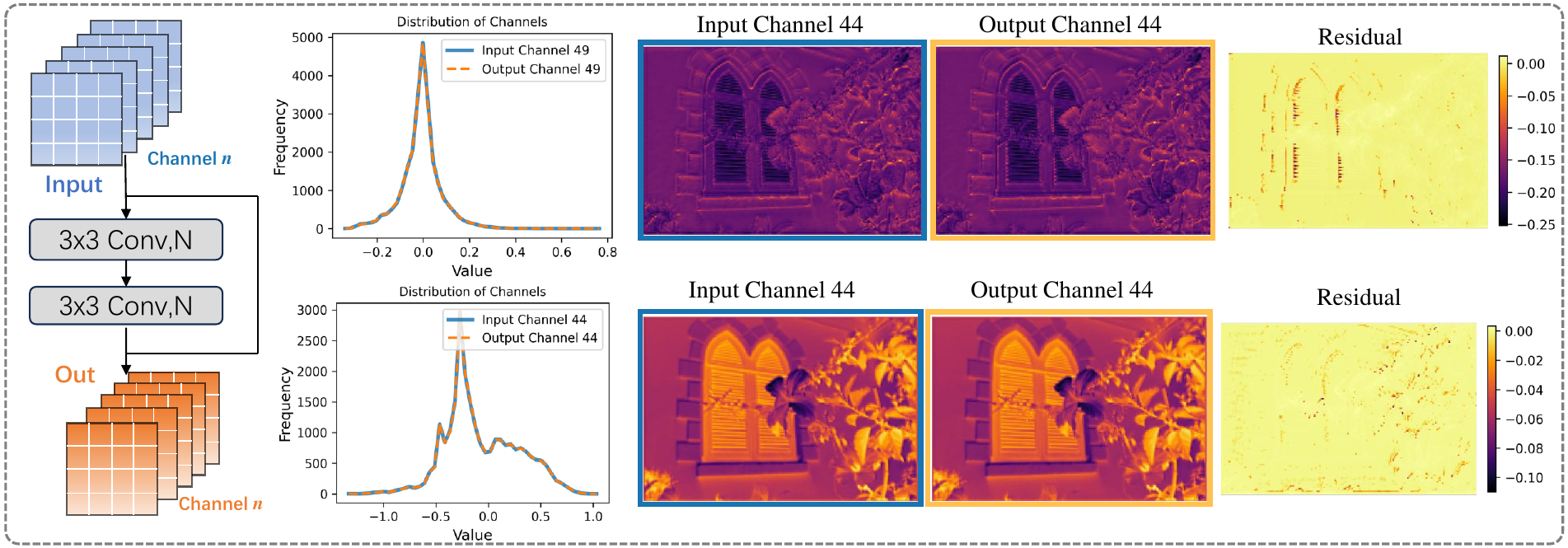}  
    \caption{\textbf{Feature redundancy in the backbone network.}  The first column  illustrates the encoder component of the Cheng2020\cite{cheng2020learned} model.
    The second column displays a histogram of channels with the same distribution. The values of these features are also the same.
    Columns three and four depict the visualization of these features.
    The fifth column highlights  the disparity between the input and output, both encapsulating edge information. 
    Our objective is to create edge features using non-parametric and less computationally demanding operations, such as shift operations.} 
    \vspace{-1em}
    \label{fig:fig-motivation}  
    \end{figure*}

We first examined the features similarity in the backbone network of Learned Image Compression (LIC). As illustrated in Figure \ref{fig:fig-motivation}, using cheng2020\cite{cheng2020learned} as an example, we observed that the features extracted by convolutional layers at different depths exhibit similarities, with the primary differences being in the representation of edges and textures. Given that shift operations can generate edge details\cite{nie2022ghostsr} and are parameter-free, they effectively reduce the number of parameters in the model\cite{wu2023fully}.\par
Inspired by these findings, we replaced the high-parameter 3$\times$3 resblocks in the network with a more efficient method (referred to as "shift"), leading to a reduction in the network's parameter load.
Additionally, we designed a grouped fusion self-attention module based on channel dimension. This approach reduces computational load through grouping and then performs pyramid feature fusion based on channel shuffle.
Owing to the extensive application of these parameter-free shifts, both the computational and parameter burden of the backbone network is substantially reduced, making it more lightweight.
Our main contributions are as follows:
\begingroup
\color{black}
\begin{itemize}

    \item We investigate feature redundancy in LIC and introduce a novel  lightweight LIC backbone design using spatial and channel shift operations. This reduces computational redundancy by efficiently utilizing similar features with free-parameter operations.
   \item We propose two key modules: the Spatial Shift Block (SSB) and the Channel Recursive Attention (CRA) module. These modules synergistically leverage shift operations in both spatial and channel dimensions to achieve efficient feature interaction, receptive field expansion, and enhanced feature representation, all while maintaining a low computational cost.
\end{itemize}
\endgroup
Our method not only achieves exceptional performance and complexity but also maintains a balance between performance and latency. Notably, ShiftLIC surpasses VVC in performance, with a computational load of merely 200KMACs/pixel, a parameter count of just 11.7M, and a GPU decoding latency of 100ms, outperforming other neural network-based methods. When evaluating the performance gain per million parameters based on our established metrics, ShiftLIC achieves $\textbf{-2.44\%/M}$, establishing a new state-of-the-art.
Furthermore, to better understand the contribution of each component of our proposed method, we have also evaluated the impact of each component.\par

The rest of the paper is structured as follows:  Section \ref{sec:related work} introduces learned image compression and Lightweight techniques for LIC. Section \ref{sec:method} provides a detailed description of our proposed method. Section \ref{sec:Experiments} presents our experimental results and performance analysis. Section \ref{sec:ablation} discusses the ablation study, evaluating the impact of different components in ShiftLIC. Finally, in the conclusion, we summarize our research findings and discuss potential future work.

\section{Related Works}
\label{sec:related work} 
In this section, we present related work in two main areas.
Firstly, we delve into learned image compression, an emerging area of research. Secondly, we discuss the  techniques for lightweight  learned image compression, which is the focus of this paper.
\vspace{-0.5em}
\subsection{Learned Image Compression}
The most effective approaches in Learned Image Compression (LIC) currently employ the variational autoencoder (VAE) structure, comprising transformation, quantization, and entropy coding stages.
\subsubsection{Transformation}
This stage effectively maps input images to compact latent representations, reducing redundancy. Generalized Divisive Normalization (GDN) initially proved effective in gaussianizing local joint statistic\cite{balle2015density}. Later, GDN was replaced by stacked residual modules in \cite{he2022elic, bao2023nonlinear}. 
To incorporate global information, non-local attention mechanisms were introduced\cite{chengtong2021tip,cheng2020learned}. Subsequently, Transformer\cite{NIPS2017_3f5ee243} have been proven to capture a more extensive range of global information\cite{qian2022entroformer,zou2022the, lu2022high, liu2023learned}. Moreover, reversible architectures have been employed within the main framework of transformation in \cite{xie2021enhanced,ma2020end}.

\subsubsection{Quantization}
Quantization discretizes continuous latent representations, but end-to-end optimization requires gradient backpropagation. Initially, additive standard uniform noise was  utilized as a soft approximation for quantization\cite{balle2017end,minnen2018joint,cheng2019learning}. Subsequent strategies include  straight-through estimators\cite{mentzer2018conditional} and annealing algorithms for approximating quantization during training\cite{Guo2021SoftTH,yang2020improving}.\par

\subsubsection{Entropy Coding}
Research on this area focuses on  accurate entropy estimation and entropy model acceleration. Hyperprior\cite{balle2018variational} models initially led the way in precise entropy estimation, followed by Gaussian Mixture Model\cite{cheng2019learning} and  Gaussian-Laplacian-Logistic Model\cite{Fu2021LearnedIC}. Subsequently, potential correlations have been modeled, such as Context model\cite{minnen2018joint}, Channel-Wise Autoregressive\cite{Channel-wise, he2022elic}, Global Context Model\cite{qian2020learning}, Cross-channel relationship model\cite{Ma2021ACC}, Causal Context Model\cite{guo_causal_2022}. 
In terms of accelerating entropy models, the main strategies involve parallel architectures, such as the Checkerboard \cite{He_2021_CVPR}, and Multi-Stage models \cite{lu2022high, Jiang2023MLIC}. \par

These advancements have enabled LIC methods to outperform the VVC standard, though at the expense of increased complexity in LIC method. Figure \ref{fig:rd-c} highlights the parameter count and computational requirements of these  state-of-the-art (SOTA) methods, underscoring the complexity challenges in deploying LIC methods in resource-limited settings.

\begin{figure}[t]
    \centering 
    \includegraphics[width=0.5\textwidth]{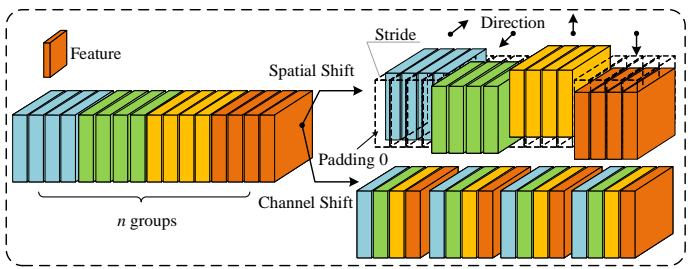}  
    \caption{\textbf{Spatial and channel shift operations.} The spatial shift involves grouping features and then shifting them in different directions. The channel shift involves rearranging the positions of the features.} 
    \label{fig:core-idea}  
    \vspace{-1em}
    \end{figure} 
\subsection{Lightweight Learned Image Compression}
Lightweight neural network are crucial for implementing models in environments with limited resources. In Learned Image Compression, the focus has primarily been on network pruning and low-bit quantization.\par
\subsubsection{Network Pruning}
Network Parameter Pruning involves identifying and eliminating redundant or less critical parameters within a neural network\cite{han2015deep, blalock2020state}. Kim \textit{et al.} proposed filter pruning and weight pruning in the decoder of LIC\cite{kim2020efficient}. Yin \textit{et al.} explored the sparsity in LIC networks and adopted an unstructured pruning method\cite{yin2022exploring}. Luo \textit{et al.} suggested structured pruning in Hyperprior component\cite{luo2022memory}. Chen \textit{et al.} determined  pruning rates for each network layer based on the layer's reliance on edge information\cite{Chen2023Pruning}. Wang \textit{et al.} increased network sparsity through a decaying $Lp$ regularization term, allowing for real-time decoding after pruning\cite{guo-hua2023evc}.

\subsubsection{Low-bit Quantization}
Low-bit Quantization involves reducing the precision of neural network weights, converting the  standard 32-bit floating-point numbers into lower bit-depth formats\cite{hubara2018quantized, qin2020binary}. 
Ball{\'{e}} \textit{et al.} used integer quantization for LIC with the aim of solving cross-platform decoding failure issues\cite{ballé2018integer}. Subsequently, He \textit{et al.} further optimized this problem using Post-Training Quantization (PTQ) techniques\cite{he2022posttraining, nagel2020up}. For model compression, Sun \textit{et al.} suggested clipping the weights' values to reduce low-bit representing error\cite{sun2020end}.  
Following this, Hong \textit{et al.} proposed layer-wise fine-tuning\cite{hong2020efficient}. Jeon \textit{et al.} suggested quantizing both weights and activations simultaneously to further make the model lighter\cite{jeon2023integer}. Sun \textit{et al.} proposed using channel splitting to minimize distortion drop\cite{sun2022qlic}, while Shi \textit{et al.} developed a channel-wisely PTQ strategy towards R-D tasks\cite{Shi2023Rate}.

\begin{figure*}[t]
    \centering 
    \includegraphics[width=\textwidth]{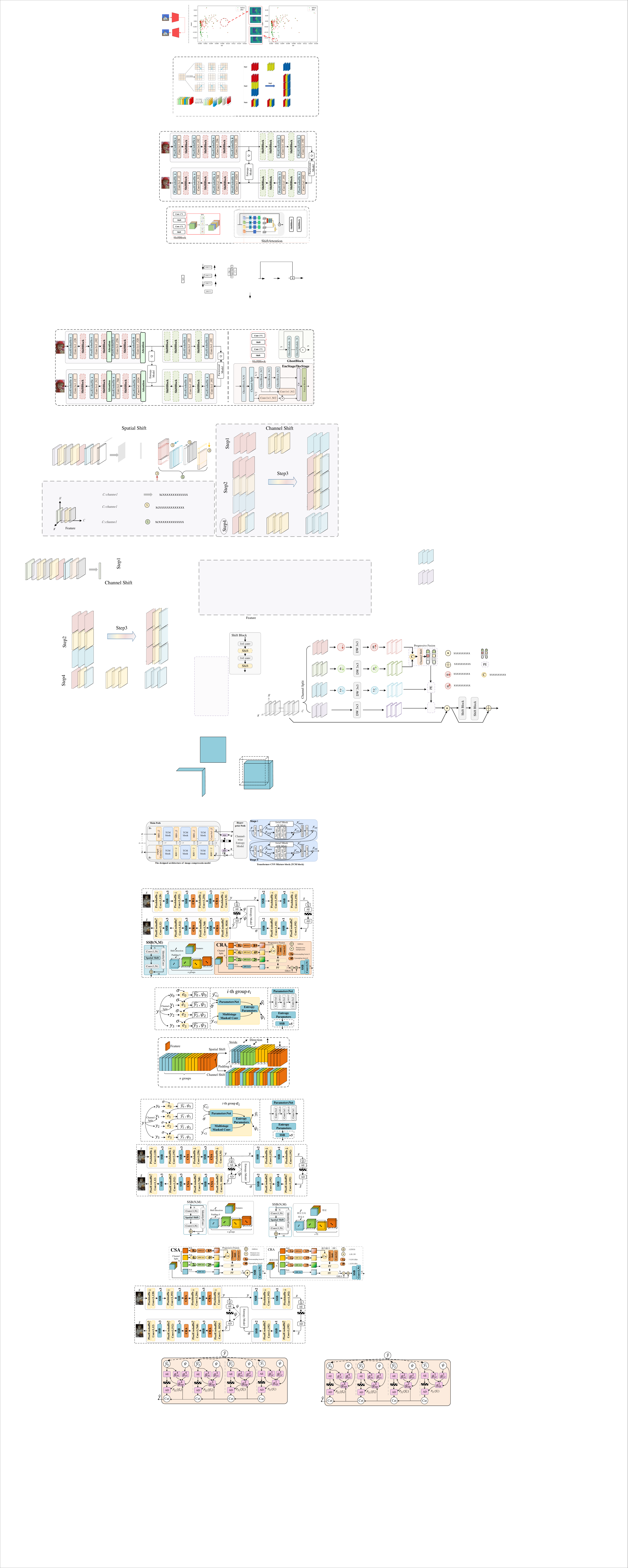}  
    \caption{
        \textbf{The proposed ShiftLIC.} Pixelshuffle is used for downsampling, and pixelunshuffle for upsampling.   `Conv(1,N)' denotes a convolution with kernel size 1 and N output channels. `SSB $\times$3' indicates that three Spatial-Shift Blocks (SSB) are used at this location. In SSB, if the number of input channels $M$ is not equal to $N$, a convolution is used in the shortcut branch for channel transformation; otherwise, a direct connection is used. In Channel Recursive Attention (CRA), DW3$\times$3 represents a depthwise convolution with a kernel size of 3. `Cat' denotes concatenation operation. \textbf{Q} represents quantization, \textbf{AE} and \textbf{AD} represent the arithmetic encoder and arithmetic decoder, respectively. Entropy Model denotes the entropy estimation module. We propose three configurations of the model: large, medium, and small. Different models use different entropy estimation modules, and the specific configurations can be seen in Table \ref{table:nic methods}.
    } 
    \vspace{-0.8em}
    \label{fig:framework}
    \end{figure*}

\begin{table}[t]
        \caption{\textbf{Notations and Full Names in the Manuscripts.}}
        \label{table:notations}
        \centering
        \small
        \renewcommand\arraystretch{0.95}
        \setlength{\tabcolsep}{8pt}{ 
        \begin{tabular}{c|c}
        \toprule
        \textbf{Full Name} &  \textbf{Abbreviation}\\ \hline
        Joint Photographic Experts Group &JPEG        \\ 
        High-Efficiency Video Coding      &HEVC          \\ 
        Better Portable Graphics         &BPG       \\ 
        Versatile Video Coding       &VVC       \\  \hline
        Variational Auto-Encoder       &VAE       \\ 
        Learned Image Compression      &LIC       \\ 
        Bjøntegaard Delta Rate      &BD-rate       \\ 
        Mean Square Error  &MSE    \\ 
        Floating-point Operation   &FLOPs  \\ 
        Peak Signal-to-Noise Ratio   &PSNR    \\ \hline        
        Spatial Shift Block      &SSB       \\ 
        Channel Recursive  Attention     &CRA        \\  
            \bottomrule
        \end{tabular}}    
    \vspace{-0.5em}
\end{table}
These methods all employ simulated quantization, whereas Le \textit{et al.} and Van \textit{et al.} focused on implementing quantized models on mobile devices, achieving real-time encoding and decoding\cite{le2022mobilecodec, van2023mobilenvc}. \par
While pruning and quantization optimize existing models, they often require specific hardware for acceleration. Lightweight algorithm  involves building efficient models from scratch, offering more flexibility across various platforms. \par
In terms of lightweight neural network architecture design, the usual strategy involves reducing the size of the convolution kernels\cite{iandola2016squeezenet} and the number of convolution groups\cite{howard2017mobilenets, zhang2018shufflenet}. 
Liu \textit{et al.} drew inspiration from transformer to design an efficient CNN backbone\cite{liu2022convnet}. EfficientNet\cite{tan2019efficientnet} was derived from the best network architecture through Neural Architecture Search (NAS)\cite{zoph2017neural}. Han \textit{et al.} designed the efficient GhostNet\cite{han2020ghostnet} based on cost-effective group convolutions. Chen \textit{et al.} combined Mobile-Net with Transformer, proposing Mobile-Former\cite{chen2022mobile}. Li \textit{et al.} utilized NAS technology in conjunction with transformer, introducing the EfficientFormer\cite{li2023rethinking} architecture.\par

These aforementioned lightweight techniques, mainly investigated for semantic tasks such as classification, have rarely been designed for LIC model. This paper aims to develop a lightweight backbone network tailored for LIC models, offering a more flexible and universal lightweight solution.

\section{Proposed Approach}
\label{sec:method}

\begingroup
\color{black}
In this section, we detail the proposed ShiftLIC. The motivation for shift operations, based on feature redundancy in LIC, is first outlined. Subsequently, the core modules, specifically the Spatial Shift Block (SSB) and Channel Recursive Attention (CRA), are introduced in detail. Finally, the network implementation is presented, demonstrating how these modules are integrated to constitute the complete ShiftLIC framework. For better understanding, notations are provided in Table \ref{table:notations}.
\endgroup

\subsection{Overview}
The redundancy in convolutional neural networks has been extensively studied, but feature redundancy within LIC algorithms has rarely been explored. Taking the Cheng2020\cite{cheng2020learned} model as an example, we analyze the feature layers of the backbone network to observe the variance in feature distribution. 
The results are shown in Figure \ref{fig:fig-motivation}. Specifically, Figure \ref{fig:fig-motivation} displays the input features of the resblock module and their output feature distribution.\par
From the figure, it is evident that the features of certain channels are remarkably similar, particularly in their edge information. This suggests a prevalent redundancy in features within LIC algorithms, leading us to question the efficiency of current feature extraction methods. Given the substantial similarities among these features and considering the high computational cost of complex feature extraction processes, we propose exploring more cost-effective and lightweight operations for feature extraction in LIC algorithms. \textbf{Is it feasible to employ `cost-effective and lightweight' operations for feature extraction in LIC algorithms?}\par

In other visual tasks, shift operations have been proven to be effective and inexpensive\cite{nie2022ghostsr, wu2023fully}. In this paper, based on shift operations, we have designed a cost-effective feature extraction module by implementing shift operations in two dimensions: spatial and channel. The overall framework of the LIC algorithm is illustrated in the Figure \ref{fig:framework}.
Next, we will introduce the details of designing shift operations in these two dimensions.\par

\subsection{Spatial-Shift Block}

\subsubsection{Spatial Shift}
\label{subsec:successive}
The spatial shift operation, as shown in Figure \ref{fig:core-idea}, is to offset the feature map. For an input feature maps $I \in R^{H \times W \times C}$,  the output feature after spatial shifting can be formulated as follows:
\begin{equation} 
O_{c,i,j}=I_{c,i+\alpha,j+\beta},
\label{eq:spatialshift1}
\end{equation}
where $O_{c,i,j}$ denotes the element value at position $\{i,j \}$ on the $c$-th feature of the output, $c$ is the channel index.  $\alpha$ and $\beta$ denotes the vertical and horizontal offsets, respectively. \par
In order to reduce the total amount of shifts, it is common to use grouped shifts, where the offsets are the same in the same group. This displacement assignment can be formulated as follows:
\begin{equation}   
\begin{aligned}    
    O^{g}_{c,i,j}=I^{g}_{c,i+\alpha_{g},j+\beta_{g}},
\end{aligned}
\end{equation}
where $O^{g} \in R^{H \times W \times C//g}$, $g$ is the group index, $C//g$ represents the result of dividing the input channels by the number of groups, emphasizing the even division of the input channels into groups, and $\alpha_{g}$ and $\beta_{g}$ denotes the vertical and horizontal offsets in the $g$-th group, respectively.\par
The spatial shift operation itself is parameterless and does not require floating-point operations (FLOPs), but the shift provides many benefits.
First, the spatial shift operation provides the information communication of the convolutional filter in the spatial dimension. Cascading a convolutional layer after a shift operation is equivalent to increasing the receptive field of that convolution.
In addition, the shift operation will facilitate the convolutional network to extract texture features, which are useful for image reconstruction. \par
The specific implementation of spatial shifting is shown in Figure \ref{fig:framework}, which refers to the method in \cite{wu2023fully}. First the channels are grouped, then the feature maps are shifted and padded with zeros. A residual bottleneck layer is defined using a 1$\times$1 convolution-shift-1$\times$1 convolution structure, named as \textbf{Spatial Shift Block}. To make the entire network algorithm more lightweight, this residual block is extensively utilized.

\subsubsection{Complexity Analysis of SSB} 
The structure of SSB is shown in Figure \ref{fig:framework}, when the number of channels of the input is not equal to the number of channels of the output, the residual connection is a $1\times1$ convolution, on the contrary, when the two are equal, it is a direct connection. For a Spatial Shift Block (SSB), $\text{Input} \in R^{H \times W \times N}$ to $\text{Output}  \in R^{H \times W \times M}$, the number of parameters is
\begin{equation}
    Params_{\text{SSB}} = \begin{cases}N^{2}+MN &M=N
        \\N^{2}+2MN &M \neq N \end{cases}.
    \label{eq:parameters1}
\end{equation}
\begingroup
\color{black}
If M does not equal N, a convolution is needed to transform N to M to ensure they can be added together. Thus, a transformation convolution parameter is added to the SSB. 
\endgroup
The total floating operations of Spatial Shift Block is
\begin{equation}
    FLOPs_{\text{SSB}} = \begin{cases} HW(N^{2}+MN) &M=N
        \\HW(N^{2}+2MN) &M \neq N \end{cases}.
    \label{eq:parameters1}
\end{equation}
We only consider the multiplication operation between float numbers when counting FLOPs.

\subsection{Channel Recursive Attention}
\subsubsection{Channel Recursive Attention}
Shifting of channels is evinced to fuse the information of channels. Following this principle, we combine the shifts in both spatial and channel dimensions to design an efficient and lightweight self-attention with the structure shown in Figure \ref{fig:framework}. 
The proposed self-attention consists of two parts: multi-scale feature extraction and regressive feature fusion.\par
In the feature extraction part, in order to reduce the complexity, grouped pyramid strategy\cite{sun2023safmn} and depth-wise convolution are employed. Specifically, for an input feature $X_I \in R^{H \times W \times C}$, first $X_I$ is sliced into $n$ groups in the channel dimension:

\begin{equation}
X_{I} \xrightarrow{split}\{X_0,\cdots,X_{n-1}\},
\label{eq:split}
\end{equation}
and then downsampling $X_n$:
\begin{equation}
    X_i=\downarrow_{{2^i}}(X_i),\ 0 \leq i\leq {n-1},
\label{eq:downsampling}
\end{equation}
where $\downarrow_{{2^i}}$ denotes downsampling and is sampled in multiples of ${2^i}$. In this way, a set of pyramid input features is obtained. Then, feature extraction is performed by convolution:
\begin{equation}
    X_i=f_{DW}(X_i),\ 0 \leq i\leq {n-1},
\label{eq:dwconv}
\end{equation}
where $f_{DW}$ denotes depth-wise convolutional layer. The algorithm is specifically being implemented with the grouping $n$ taken to be 4 and $f_{DW}$ is employed $3\times3$ depth-wise convolutional layer.
Next, these grouped features need to be fused. These features are first upsampled and restored to the same size.
\begin{equation}
    X_i=\uparrow_{{2^i}}(X_i),\ 0 \leq i\leq {n-1},
\label{eq:upsampling}
\end{equation}
where $\uparrow_{{2^i}}$ denotes upsampling and is sampled in multiples of ${2^i}$.  
The feature fusion part is shown in Figure \ref{fig:framework}, where the features are channel-wise concatenated, then go through a channel shuffle, and then through a convolutional layer, and the whole process can be formulated as follows:
\begin{equation}   
    \begin{aligned}
    Y_1 &= F_{conv}(\text{ChannelShuffle}(\text{Concat}(X_0, X_1))), 
    \end{aligned}
\end{equation}
where $\text{Concat}$ denotes a concatenation operation along the channel dimension, $Y_1$ is the intermediate feature and $F_{conv}$ employs a $ 1\times 1 $ convolution. Denote the above Feature Shuffle Fusion process as $f_\text{FSF}(X_0, X_1)$. 
The final recursively fused features can be formulated as follows: 
\begin{equation}   
    \begin{aligned}
    Y &= f_\text{FSF}(f_\text{FSF}(f_\text{FSF}(X_0, X_1), X_2),X_3)
    \end{aligned}
\end{equation}

Finally, the resulting $Y$ undergoes a nonlinear funtion GELU, which serves as the feature map for self-attention, and is multiplied element-wise with the input $X_I$. Then, residual skip connections are applied. This process can be written as:
\begin{equation}
    {X}_{O} =\text{GELU}(Y)\odot X_I + X_I,
\end{equation}
where $\odot$ denotes the element-wise multiplication, $X_O \in R^{H \times W \times C} $ is the output feature. $X_O$ is with global information, but local features are also needed. In order to supplement the local information, the SSB is employed again and combined with the channel dimensioning operation to enhance the local feature extraction capability. The final algorithm is shown in Figure \ref{fig:framework}.\par
\subsubsection{Complexity Analysis of CRA}
For a Channel Recursive Attention (CRA), $\text{Input} \in R^{H \times W \times N}$ to $\text{Output}  \in R^{H \times W \times N}$, the number of parameters is
\begin{equation}
    Params_{\text{CRA}} = 9N + \frac{39}{8}N^2,
    \label{eq:parameters2}
\end{equation}
The total floating operations of Channel Recursive Attention is
\begin{equation}
    FLOPs_{\text{CRA}} =HW(\frac{765}{256}N +7\frac{13}{16}N^2).
    \label{eq:parameters2}
\end{equation}
\par
\begingroup
\color{black}
Because the compression algorithm uses multiple size sampling operations, feature map sizes vary. Convolutional layer computation is tied to input size; thus, CRA module placement within the algorithm impacts computational complexity. Subsequent experiments will analyze how the number and location of CRA modules affect algorithm performance and complexity.
\endgroup
\par

\begin{figure*}[t]
        \centering
        \begin{subfigure}[b]{0.49\linewidth}
          \includegraphics[width=\linewidth]{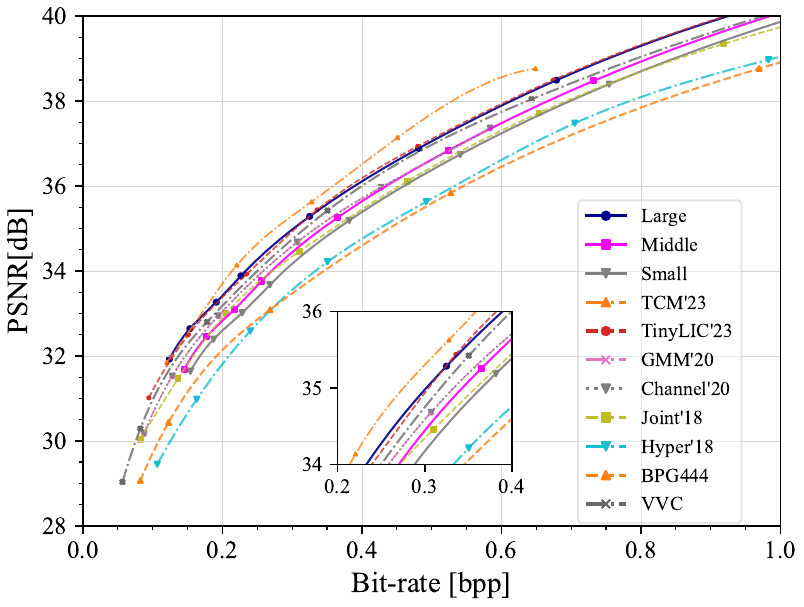}
          \caption{CLIC Valid/MSE optimized}
          \label{fig:kodak-mse}
        \end{subfigure}
        \vspace{0em}
        \begin{subfigure}[b]{0.49\linewidth}
          \includegraphics[width=\linewidth]{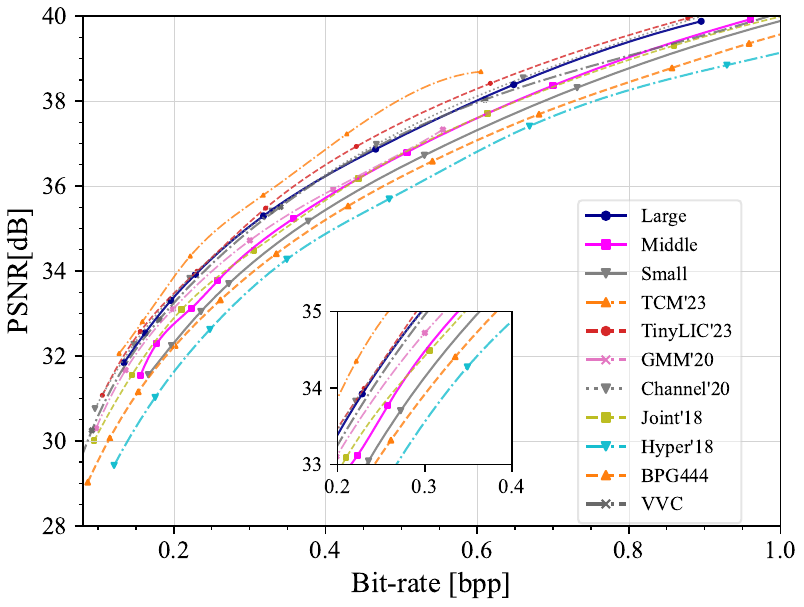}
          \caption{Tecnick/MSE optimized}
          \label{fig:kodak-msssim}
        \end{subfigure}
        \vspace{0em} 
        \begin{subfigure}[b]{0.485\linewidth}
          \includegraphics[width=\linewidth]{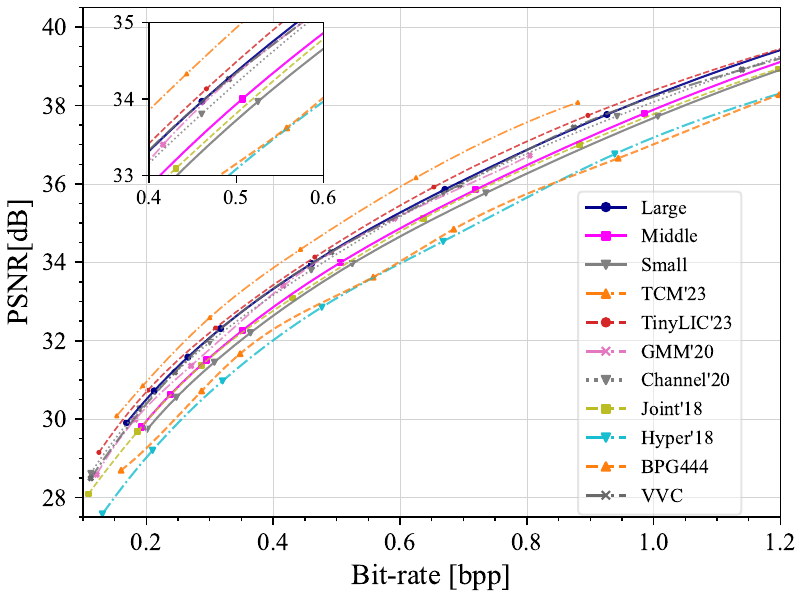}
          \caption{Kodak/MSE optimized}
          \label{fig:tecknick-mse}
        \end{subfigure}
        \vspace{0em} 
        \begin{subfigure}[b]{0.495\linewidth}
          \includegraphics[width=\linewidth]{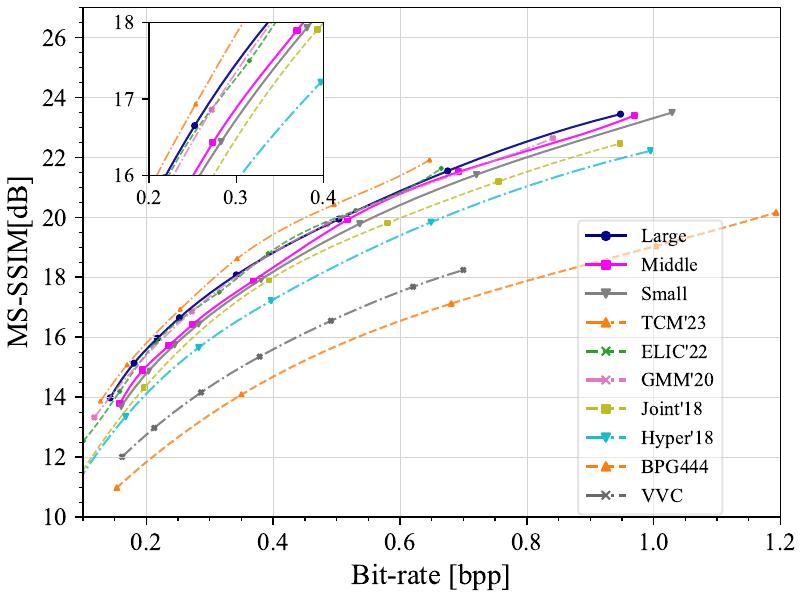}
          \caption{Kodak/MS-SSIM optimized}
          \label{fig:clic-mse}
        \end{subfigure}
        \caption{\textbf{Comparisons of RD curves with other methods on the Kodak, Tecnick, and CLIC-valid test datasets.}  (d) show the results with MS-SSIM as a reconstruction loss. For a clear illustration, the MS-SSIM values are converted into decibels  $-10log_{10}(1-d)$, where $d$ refers to the MS-SSIM value. The anchor for calculating the BD-rate gain is BPG with YUV444 configuration. For detailed quantitative results of each method, please refer to Table \ref{table:quantitative}.}
        \label{fig:RD}
\end{figure*}

\begin{figure*}[t]
    \color{blue}    
    \begin{subfigure}[b]{0.485\linewidth}
      \includegraphics[width=\linewidth]{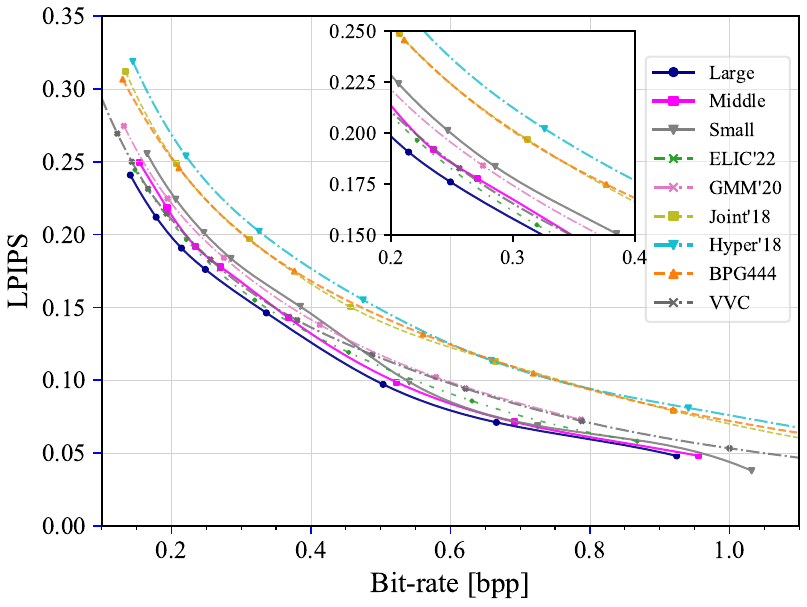}
      \caption{Performance under LPIPS metric}
      \label{fig:jpegai-lpips}
    \end{subfigure}
    \vspace{0em} 
    \begin{subfigure}[b]{0.495\linewidth}
      \includegraphics[width=\linewidth]{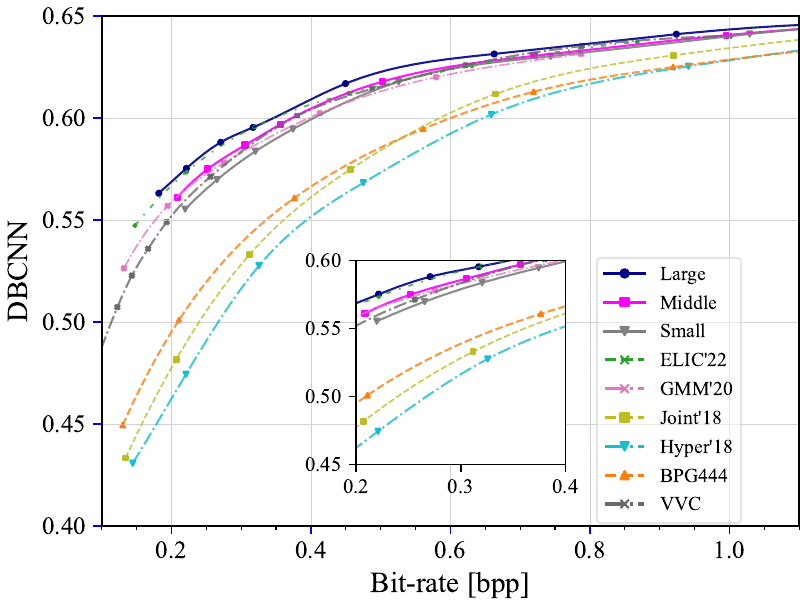}
      \caption{Performance under DBCNN metric}
      \label{fig:jpegai-dbcnn}
    \end{subfigure}
    \caption{{\textbf{Comparisons of RD curves with other methods on the JPEG-AI test dataset.}  (a) and (b) show the results with LPIPS\cite{zhang2018unreasonable} and DBCNN\cite{zhang2020blind} as distortion metrics, respectively.}}
    \label{fig:RD2}
\end{figure*}

\begin{table}[t]
    \caption{\textbf{Complexity Comparison.} Only multiplication is counted. Due to the merging of all calculation results, numbers such as 765/256 are generated.}
    \vspace{-0.3em}
    \label{table:nic methods}
    \centering
    \small
    \renewcommand\arraystretch{1.2}
    \setlength{\tabcolsep}{5pt}{ 
    \begin{tabular}{c|cc} 
    \toprule
    \multirow{1}{*}{\textbf{Methods}}   &Params   &FLOPs \\ \hline
    3x3 resblock  &9$M^{2}+18MN$    &$HW(9M^{2}+18MN)$   \\ 
    SSB    & $N^{2}+2MN$   &$HW(N^{2}+2MN)$    \\ \hline

    Attention  &$20\frac{1}{2}N^2$    &$HW(20\frac{1}{2}N^2+N)$   \\ 
    Nonlocal  &$22\frac{1}{2}N^2$   &$HW(22\frac{1}{2}N^2+N+HWN)$  \\ 
    CRA    &$4\frac{7}{8}N^2+9N$    &$HW(7\frac{13}{16}N^2 + \frac{765}{256}N)$    \\ 

    \bottomrule
    \end{tabular}}    
\end{table}

\subsection{Network Implementation}
The overall network structure is shown in Figure \ref{fig:framework}. The LIC backbone network has been redesigned based on the two lightweight modules mentioned above. The main improvements are as follows:
\begin{itemize}
    \item \textbf{Pixel Shuffle Sampling.}  The convolution-based sampling is replaced by the pixel shuffle strategy to minimize the loss due to sampling. At the same time, in order to reduce the complexity, $1\times1$ convolution is used to reduce the number of channels in the output of pixel shuffle.
    \item \textbf{Spatial-Shift Block.}  Due to the low number of parameters and computational requirements of the SSB unit, it is extensively used within the network. In the main encoder/decoder, it serves as a non-linear transformation module, with multiple shift blocks stacked in each section to ensure algorithm performance. In the hyper encoder/decoder, the number of shift blocks is reduced.
    \item \textbf{Channel Recursive Attention.}   The attention module is relatively complex, and its position within the algorithm structure determines the computational complexity. Therefore, it is embedded in the middle and end parts of the main encoder/decoder.
    \item \textbf{Entropy Model.}  This paper does not involve improvements in the structure of the entropy model, but this module determines the encoding and decoding time. To validate the efficiency of our proposed backbone LIC algorithm, we employed two types of entropy model algorithms, named  Hyperprior\cite{balle2018variational} and Multistage Context Model\cite{lu2022high}.
\end{itemize}
\par
In the following sections, we will experimentally verify the effectiveness of Shift-LIC and its adaptability to different entropy modeling algorithms.

\begin{table}[t]
    \caption{Specific configurations for three different scale models. $M$ denotes the number of channels in the entropy bottleneck layer.}
    \vspace{-0.3em}
    \label{table:nic methods}
    \centering
    \normalsize
    \renewcommand\arraystretch{1.2}
    \setlength{\tabcolsep}{8pt}{ 
    \begin{tabular}{c|ccc} 
    \toprule
    \multirow{2}{*}{\textbf{Methods}} &  \multicolumn{3}{c}{\textbf{Key Components}}  \\ \cline{2-4} 
                          &Attention   &Entropy Model  &$M$ \\ \hline
           Small  &None     & Hyperprior\cite{balle2018variational}  &320\\ 
           Medium    &CRA   & Hyperprior\cite{balle2018variational}   &320 \\ 
           Large  &CRA   & Multistage Context\cite{lu2022high}  &320 \\ 
        \bottomrule
    \end{tabular}}    
\end{table}

\begin{table*}[t]
    \vspace{1em}
    \caption{\textbf{Quantitative evaluation results.}  This involved a comprehensive assessment of the performance, complexity, and encoding/decoding latency on GPU and CPU for different methods.
    Results for the three datasets are based on their original publications. The notation {\textbackslash{}} indicates that the original study did not test on that dataset. FLOPs and the number of parameters were uniformly measured using \textit{ptflops}, with the test size being 768$\times$512. Encoding and decoding times were tested on the same device, with the table showing average values for the Kodak dataset. VTM12.1 and BPG were tested using their default configurations. \textbf{The BDrate/FLOPs metric represents the performance gain per MACs/pixel.} \textcolor{red}{Red} indicates the best performance per column, and \textcolor{blue}{Blue} indicates the worst.}
    \vspace{-0.5em}
    \label{table:quantitative}
    \centering
    \renewcommand\arraystretch{1.1}
    \resizebox{18.5cm}{!}{
    \setlength{\tabcolsep}{2pt}
    \normalsize
    \begin{tabular}{cccccccccccc}
    \toprule
    \multirow{2}{*}{Method} & \multicolumn{1}{c}{CLIC$\downarrow$}  & \multicolumn{1}{c}{Tecnick$\downarrow$}  &\multicolumn{2}{c}{Kodak$\downarrow$}      
    
    &\multicolumn{1}{c}{\multirow{2}{*}{\begin{tabular}[c]{@{}c@{}}FLOPs$\downarrow$\\ (KMACs/pixel)\end{tabular}}}  
    &\multicolumn{1}{c}{\multirow{2}{*}{\begin{tabular}[c]{@{}c@{}}Para$\downarrow$\\ ($M$)\end{tabular}}} 
    &\multicolumn{1}{c}{\multirow{2}{*}{\begin{tabular}[c]{@{}c@{}}BDrate/FLOPs$\downarrow$\\ $\%$/(MACs/pixel)\end{tabular}}}  
    & \multicolumn{2}{c}{GPU Latency(ms)$\downarrow$} & \multicolumn{2}{c}{CPU Latency(ms)$\downarrow$}  \\  \cmidrule(r){2-2} \cmidrule(r){3-3} \cmidrule(r){4-5} \cmidrule(r){9-10}    \cmidrule(r){11-12} 
    &MSE &MSE &MSE &MS-SSIM   & & & &Enc.T  &Dec.T  &Enc.T  &Dec.T   \\ \hline \hline

    BPG  &0 &0 &0 &0 &\textbackslash{}  &\textbackslash{} &\textbackslash{} &\textbackslash{} &\textbackslash{}  &548 &\textcolor{red}{\underline{127}} \\ 
    VTM12.1 &-26.61 &-20.27 &-20.22 &-19.47  &\textbackslash{}  &\textbackslash{} &\textbackslash{}  &\textbackslash{} &\textbackslash{}  &\textcolor{blue}{\underline{74228}} &153 \\ \hline

    Hyper'18\cite{balle2018variational} &\textcolor{blue}{\underline{0.43}} &\textcolor{blue}{\underline{9.81}}  &\textcolor{blue}{\underline{2.89}} &\textcolor{blue}{\underline{-43.19}}  &418.09    & 11.82 &\textcolor{blue}{\underline{6.91}} & \textcolor{red}{\underline{42}}  & \textcolor{red}{\underline{48}}  &\textcolor{red}{\underline{225}}  & 279  \\ 
    Joint'18\cite{minnen2018joint} &-18.50 &-11.77 &-12.36 &-48.08  &449.82   & 25.5 &-27.47  &2374    & 5741    & 2807     & \textcolor{blue}{\underline{7264}} \\ 
   Channel'20\cite{Channel-wise} &-23.30 &-22.77 &-19.79 &\textbackslash{}  &610.65   & 68.39 &-32.41 & 66     & 90     & 422      & 467 \\ 
   GMM'20\cite{cheng2020learned} &-23.31 &-17.74 &-19.48 &-58.15  &1027.22    & 29.63 &-18.96 & 2363    & 5696    & 2190     & 6450  \\
   INC'21\cite{xie2021enhanced} &-28.26 &-23.83 &-21.42 & -57.88  &1038.20    & 50.03 &-20.63 &\textcolor{blue}{\underline{2801}}    & \textcolor{blue}{\underline{6859}}    & 2354     & 5740 \\ 
   SFT'22\cite{zou2022the} &-29.71 &\textbackslash{} &-24.55 &-60.73  &508.42   & 99.86 &-48.28 & 91    & 122     & 592      & 706 \\     
    ELIC'22\cite{he2022elic}  &\textbackslash{} &\textbackslash{} &-26.88 &-57.58  &833.12   & 33.79 &-32.26 & 209     & 128     & 660      & 608 \\        
    TCM'23\cite{liu2023learned}  &\textcolor{red}{\underline{-35.25}} &\textcolor{red}{\underline{-31.19}} &\textcolor{red}{\underline{-30.09}} &\textcolor{red}{\underline{-61.78}} &\textcolor{blue}{\underline{3843.68}}  &\textcolor{blue}{\underline{121.88}} &-7.82 &157     & 175     & 2805     & 3070 \\  \hline  
    TinyLIC'23\cite{lu2022high} &-29.66 &-24.43 &-23.86 &\textbackslash{}  &491.02   & 28.34 &-48.59  &102  & 118 & \textbackslash{} & \textbackslash{} \\ 
    EVC'23\cite{guo-hua2023evc} &\textbackslash{} &\textbackslash{} &-20.62 &\textbackslash{}  &557.86  &17.38 &-36.96  &190  &26 &1475 & 786 \\ 
       
    \hline

    Small &-15.43 &-4.71  &-9.55 &-52.24  &114.13  & \textcolor{red}{\underline{3.97}}  &-84.81  & 44     & 50   & 343  &342 \\ 
    Mid   &-20.12 &-11.54 &-13.28 &-54.09  &173.33   & 5.79  &-76.5  & 46    & 53   & 439   &438  \\
    Large &-29.21 &-21.06 &-20.73 &-58.78  &197.45   &11.65 &\textcolor{red}{\underline{-102.56}}   & 78   & 101  &514   &531   \\   \hline
    
    \bottomrule
    \end{tabular}
    }
\end{table*}

\section{Experiments}
\label{sec:Experiments}
\subsection{Experiment Setup}
\label{subsec:e_setup}
\textbf{Datasets.} The CLIC\footnote{\url{http://compression.cc/tasks}} dataset is used for training, where the samples is randomly cropped into 247,000 images of $256 \times 256$ patches. For test, three datasets are used, including Kodak datasets\footnote{\url{https://r0k.us/graphics/kodak/}} with 24 images at size of $768 \times 512$, Tecnick dataset\footnote{\url{https://sourceforge.net/projects/testimages/files/OLD/OLD_SAMPLING/}} with 100 high quality images size of $1200 \times 1200$ and CLIC Professional Validation dataset\footnotemark[1] with 41 2k resolution images. The average results on each dataset are computed for comparing the performance fairly. \par
\textbf{Training details.}
We train each model for 100 epochs using the Adam optimizer and a batch size is 16. The learning rate is initially set to $10^{-4}$, and then reduced to $5 \times 10^{-5}$ at the 40th epochs, following by a drop to $1 \times 10^{-5}$ at 80th epochs. The threshold of gradient clipping is 1 to stable training.
Loss function is defined as $L = R + \lambda \cdot D$, where Distortion term $D$ is measured by Mean Square Error (MSE)  and multi-scale structural similarity (MS-SSIM) between original images and reconstructed images. When $D$ is MSE, the Lagrange multiplier $\lambda$  is chosen from the set \{0.0035, 0.005, 0.0067, 0.0130, 0.0250, 0.050, 0.100\}. In terms of MS-SSIM, $\lambda \in$ \{5.0, 6.51, 8.73, 16.64, 31.73, 60.50, 140\}.\par
\textbf{Model Settings.}
The structure of ShiftLIC is shown in Figure \ref{fig:framework}. Based on different entropy models and attention modules, three scales of models are configured, with the configurations of each model detailed in Table \ref{table:nic methods}. ShiftLIC is implemented using the open-source library CompressAI. \par
In the main encoder/decoder, the number of Spatial-Shift Blocks (SSB) is set to 3, while in the hyper encoder/decoder, it is set to 2 and 1 respectively. In the SSB, the number of groups is set to 4, with shifts occurring along four diagonal directions: bottom-left, top-left, top-right, and bottom-right, with a shift step length of 1. In the Channel Recursive Attention (CRA), the channel grouping is set to 4, and the number of channel shuffle groups is 8.

\begin{figure*}[t]
        \centering
        \begin{subfigure}[b]{0.95\linewidth}
          \includegraphics[width=\linewidth]{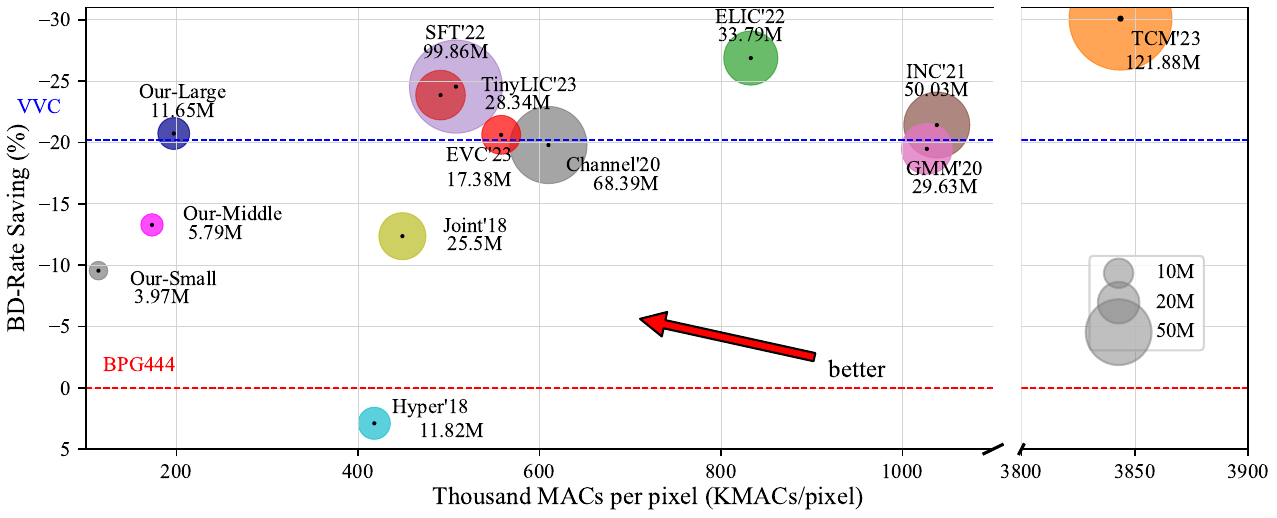}
          \caption{Kodak/MSE optimized}
          \label{fig:rd-c-all}
        \end{subfigure}
        \begin{subfigure}[b]{0.49\linewidth}
            \includegraphics[width=\linewidth]{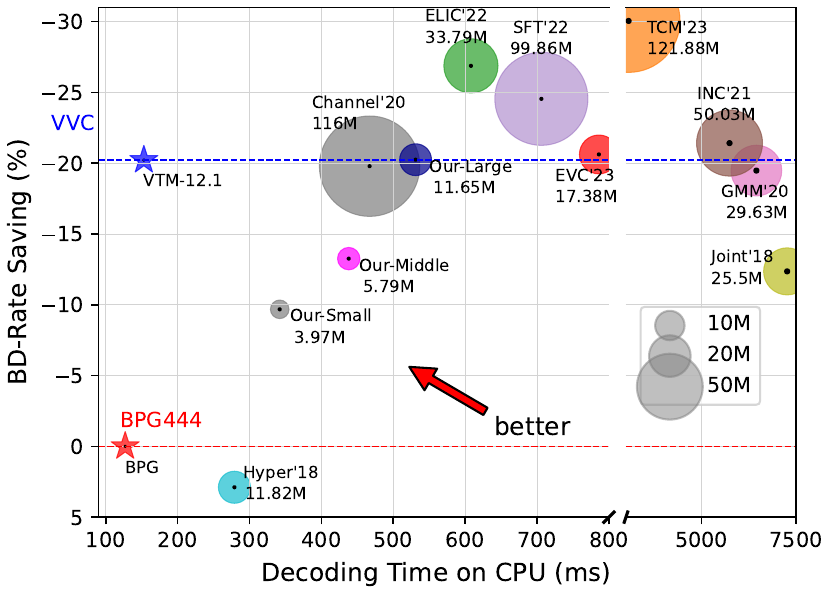}
            \caption{The average CPU decoding time on the Kodak.}
            \label{fig:rd-t-dec-cpu}
          \end{subfigure}
        \begin{subfigure}[b]{0.49\linewidth}
          \includegraphics[width=\linewidth]{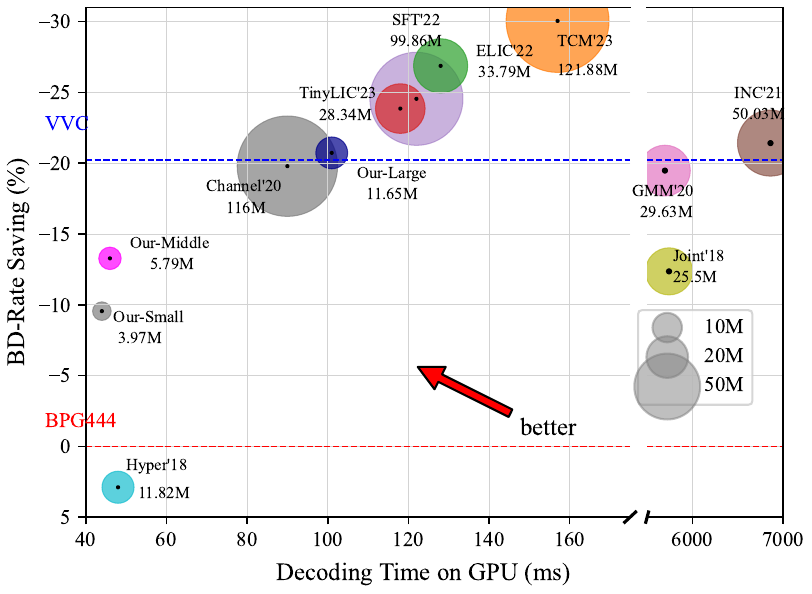}
          \caption{The average GPU decoding time on the Kodak.}
          \label{fig:rd-t-dec}
        \end{subfigure}
        \vspace{0em}
        \caption{ \textbf{Performance versus Complexity/CPU Latency/GPU Latency.} R-D performance is quantified using BD-rate, which is anchored by BPG.  All latency were measured on a unified platform, and the average latency of 24 Kodak test images were used. Thanks to years of optimization, traditional methods perform well on CPU. Among all neural network based methods, our methods achieve an exceptionally good balance between performance and latency.}
        \label{fig:decoding-time}
\end{figure*}

\textbf{Evaluation details.}
We used the BPG with YUV444 configuration as the anchor to derive BD-rate gains, and the VVC Intra reference software is VTM-12.1. For a fair comparison, we try our best to ensure a similar bitrate range across different approaches for BD-rate computation.\par

\subsection{Rate-distortion Performance}
Two distortion metrics, Mean Squared Error (MSE) and Multi-Scale Structural Similarity (MS-SSIM), are used as the reconstruction loss terms. MSE represents signal fidelity, while MS-SSIM indicates perceived quality.
All results are the average values over the test dataset. The Bjøntegaard Delta (BD) rate relative to BPG444 is used as a performance evaluation metric. The BD rate is defined as the average bit rate savings relative to the anchor method while maintaining the same quality metrics. Next, we will discuss the rate-distortion performance under these two distortion objectives.\par
\textbf{R-D Performance optimized with MSE.} 
Figures \ref{fig:RD}-(a), \ref{fig:RD}-(c), and \ref{fig:RD}-(d) display the R-D curves for the MSE as the optimization metric. Our models were compared with the state-of-the-art learned image compression algorithms. For a fair comparison, all methods in each figure use the same optimization objective. \par
Figure \ref{fig:RD}-(c) shows the test results on the Kodak dataset. Our large model approach achieved performance comparable to VVC, with a BD rate of -20.25\% relative to the anchor BPG, while VVC achieved -20.22\%.
On the high-resolution CLIC and Tecnick datasets, our performance even surpassed that of VVC. For example, relative to the anchor, the large model achieved a -28.21\% BD rate on the CLIC validation set, compared to VVC's -26.61\%. Similarly, on Tecnick, the large model achieved a -20.70\% BD rate, compared to VVC's -20.27\%. The reason for better performance at high resolution is that, compared to the VVC method, LIC approaches do not require block processing of the image, which is advantageous for capturing global information.
\par
Furthermore, compared to other SOTA neural network-based compression algorithms, the performance of our method is also very competitive. 
For more quantitative results, please refer to Table \ref{table:quantitative}.\par

\textbf{R-D Performance optimized with MS-SSIM.} 
As shown in Figure \ref{fig:RD}-(d), when visual quality is used as the optimization metric for the model, learned-based methods demonstrate a greater advantage compared to VVC. On the Kodak dataset, our method achieved a -58.72\% BD rate relative to the anchor, while VVC only achieved a -19.47\% BD rate. This is because learned-based methods are better at capturing the texture information of images, which is more beneficial for enhancing visual quality. Compared to other neural network-based methods, our large model maintains competitive performance with other state-of-the-art (SOTA) approaches. For more quantitative results, please refer to Table \ref{table:quantitative}.\par
\begin{table*}[t]
    \vspace{1em}
    \caption{\textbf{Ablation Studies of Proposed Components.}  The mark $\times$ denotes the removal of the component. the anchor of the BD-rate is the full model, i.e., case6. Smaller values mean better performance. FLOPs and parameter counts are measured uniformly using \textit{ptflops} with a test size of 256 $\times$ 256. The latency is averaged over the Kodak dataset. The difference in Latency is milliseconds.}
    \vspace{-0.5em}
    \label{table:ablation_all}  
    \centering
    
    \renewcommand\arraystretch{1.2}
    \setlength{\tabcolsep}{3.5pt}{
    \normalsize
    \begin{tabular}{cccccccccc}
    \toprule
    \multirow{2}{*}{Method} & \multicolumn{2}{c}{SSB}  & \multicolumn{2}{c}{Attention} &\multirow{2}{*}{BD-rate(\%)} & \multirow{2}{*}{Para($M$)}  &\multicolumn{1}{c}{\multirow{2}{*}{\begin{tabular}[c]{@{}c@{}}FLOPs\\ (KMAC/pixels)\end{tabular}}}     & \multicolumn{2}{c}{GPU Latency(ms)}  \\  \cmidrule(r){2-3} \cmidrule(r){4-5}  \cmidrule(r){9-10} 
    &Spatial-Shift &Following 1$\times$1  &Channel Recursive &SSB & & & &Enc.T  &Dec.T      \\ \hline  \hline 
   
    case1 &$\times$  &$\checkmark$ &\checkmark &\checkmark &6.23\% &5.79 &173.33 &51 &57  \\ 
    case2 &\checkmark  &$\times$ &\checkmark &\checkmark &4.07\% &4.58 &121.01 &46 &53\\    
    case3 &\checkmark  &\checkmark  &$\times$ &\checkmark &1.24\% &5.79 &173.33  &52 &60\\

    case4 &\checkmark  &\checkmark  &\checkmark  &$\times$ &3.35\% &4.35 &126.34  &52 &58\\
    case5 &\checkmark  &\checkmark &$\times$ &$\times$ &3.51\% &3.97 &114.14 &45 &51\\ 
    case6 &\checkmark  &\checkmark  &\checkmark  &\checkmark &Anchor(0\%) &5.79 &173.33 &48 &56\\             
    \bottomrule
    \end{tabular}
    }
\end{table*}

\begingroup
\color{black}
\textbf{R-D Performance under perceptual metrics.}
To further illustrate the effectiveness of the proposed method in perceptual quality, Figure \ref{fig:RD2} evaluates two perceptual metrics: LPIPS\cite{zhang2018unreasonable} and DBCNN\cite{zhang2020blind}, used as measures of distortion. LPIPS serves as a full-reference evaluation metric, whereas DBCNN is a no-reference evaluation metric. Figure \ref{fig:RD2}  demonstrates that the proposed large model configuration method maintains strong performance across these metrics, underscoring the advantages of our approach in improving visual quality.
\endgroup

\subsection{Latency Cost}
Encoding and decoding latency are crucial metrics for evaluating codecs. We assessed the encoding and decoding latency of various algorithms on a unified platform, where encoding latency refers to the time from inputting an image to generating an encoded stream, and decoding latency refers to the time required to reconstruct an image from the encoded stream. The test machine was configured with an Intel Xeon 4310 CPU, NVIDIA A800 GPU, and 256 GB RAM. \par
Traditional standards were tested using the CPU, as they do not support GPU acceleration. To ensure a fair comparison, both CPU and GPU times were measured, and the average was taken over 24 images from the Kodak test set.
The tested code was sourced from the original papers or open-source implementations. \par

GPU latency, as illustrated in Figure \ref{fig:decoding-time}-(a) reveals that our model methods of all three scales have relatively lower encoding and decoding latencies compared to other LIC methods. For example, in terms of latency, our small and medium models are the best performers. Even the latency of our large model is controlled within 100ms when compared to other SOTA methods like TCM'23\cite{liu2023learned}.\par
CPU latency, as shown in Figure \ref{fig:decoding-time}-(b), demonstrates that traditional methods have achieved excellent decoding times due to years of optimization. Compared to other LIC methods, our models also achieve outstanding CPU decoding times.
Considering the overall R-D performance of the models, our method achieves an excellent trade-off between R-D performance and latency.
 
\subsection{Complexity Analysis}
\par \textbf{Complexity Analysis of Backbone} \: As shown in Table \ref{table:quantitative}, we compared the number of parameters and computational complexity of several LIC algorithms. Our three models are the most competitive in terms of performance-complexity trade-off.\par
For instance, the large model achieves a BD rate of -20.25\% relative to the anchor BPG444 on Kodak, comparable to some SOTA methods like Channel'20\cite{Channel-wise}, GMM'20\cite{cheng2020learned}, and INC'21\cite{xie2021enhanced}.
However, the parameter count of the large model is approximately 1/6th of Channel'20 model, 1/3rd of GMM'20 model, and 1/5th of INC'21 model, while its computational complexity is about 1/3rd of Channel'20 model, 1/5th of GMM'20 model, and 1/5th of INC'21 model. This indicates that our method has a superior performance-complexity frontier.
Our medium model surpasses the performance of the Joint'18\cite{minnen2018joint} model (e.g., -13.26\% vs -12.36\% on Kodak), but with only about 1/5th the number of parameters and about 1/3rd the computational complexity of the Joint'18 model.
Furthermore, our small model's performance is close to the Joint'18 model, but with only about 1/6th the number of parameters and about 1/4th the computational complexity.\par

\par \textbf{Complexity Analysis of Entropy Models.} \: Table \ref{table:entropy complexity} shows the complexity of different entropy models. Our method, MCM*, employs SSB as an entropy parameter estimation network. Compared to MCM, it reduces both computational load and the number of parameters. In contrast to other entropy models, MCM* strikes a balance between parameter count, computational complexity, and time complexity. \par
\vspace{1em}
In summary, our method significantly advances the rate-distortion-computation frontier and represents the best current option for performance-complexity trade-offs, with potential for mobile deployment.

\begin{table}[t]
    \caption{Comparison of the complexity of different entropy models.}
    \vspace{-0.3em}
    \label{table:entropy complexity}
    \centering
    \normalsize
    \renewcommand\arraystretch{1.2}
    \setlength{\tabcolsep}{6pt}{ 
    \begin{tabular}{c|ccc} 
    \toprule
    \textbf{Methods}  &FLOPs(G)   &Para(M)  &Complexity \\ \hline
           Hyper\cite{balle2018variational}  &0.65  &4.81  &$\mathcal{O}(1)$\\ 
           Autoregressive\cite{minnen2018joint}    &1.09   &8.35  &$\mathcal{O}(HW)$ \\ 
           ChannelWise\cite{Channel-wise}   &13.27   &61.38   &$\mathcal{O}(10)$ \\
           ELIC\cite{he2022elic}  &5.01   &19.12   &$\mathcal{O}(10)$ \\
           MCM\cite{lu2022high} &3.2   &18.74  &$\mathcal{O}(9)$ \\
           MCM*(Ours) &1.74   &7.14  &$\mathcal{O}(9)$ \\
        \bottomrule
    \end{tabular}}    
\end{table}

\subsection{Qualitative Visualization}
\begin{figure*}[t]
    \centering 
    \includegraphics[width=\textwidth]{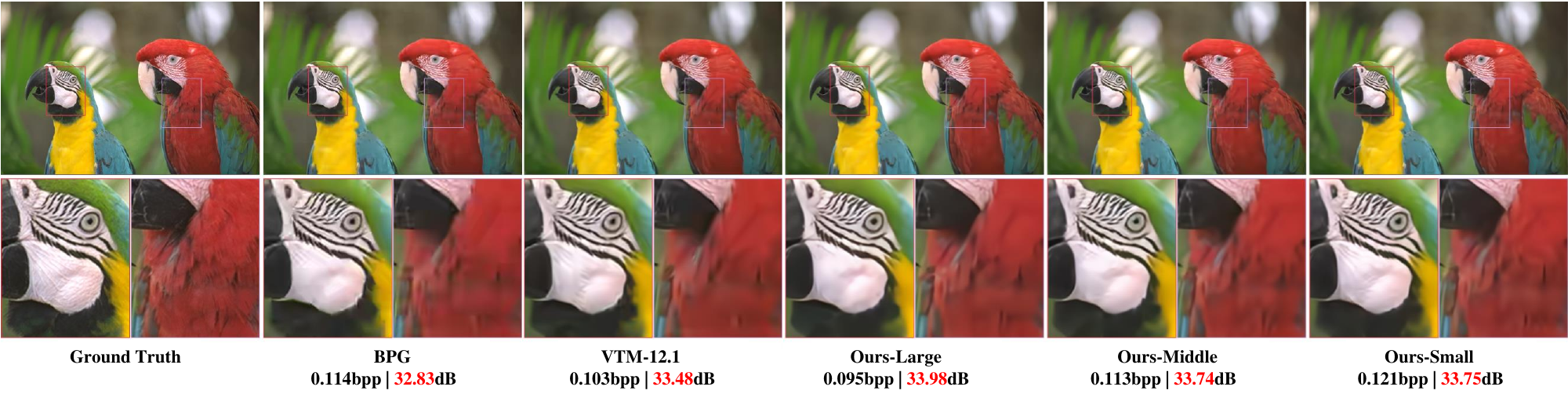}
    \caption{\textbf{Qualitative Visualization.} The reconstructions and close-ups of images encoded by BPG, VVC, and our method at around 0.1 bpp are shown. Annotations under the cropped images indicate the bpp/PSNR (dB) for the entire image. In the reconstructed images of our Large model, more "feathers" can be clearly seen, indicating that our method is capable of capturing more texture information.} 
    \label{fig:qua-vis1}  
    \end{figure*}

Figure \ref{fig:qua-vis1} visually demonstrates the quality of the reconstructed images. To ensure a fair comparison, the bpp (bits per pixel) values of all reconstructed images are kept close. Compared to BPG, our proposed method significantly enhances visual quality, and in comparison to VVC, our method exhibits superior performance in capturing certain texture details. For more results on visual quality, please refer to our open-source resource repository. \par

\begin{figure}[t]
    \centering 
    \includegraphics[width=0.40\textwidth]{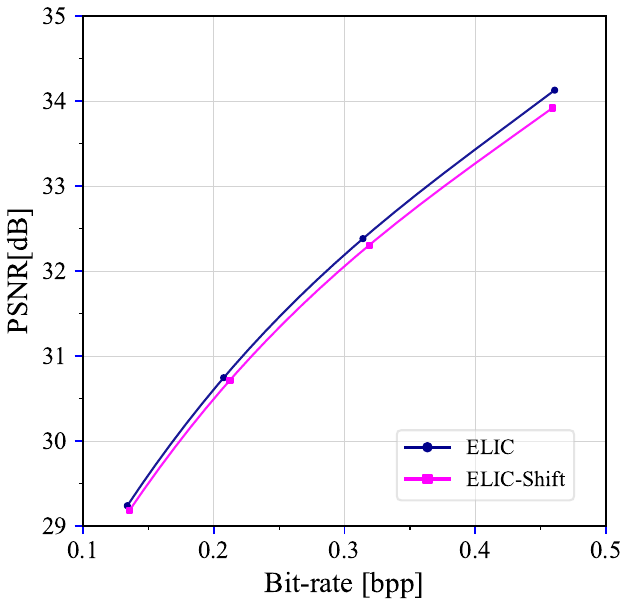}  
    \vspace{-0.5em}
    \caption{RD Performance of the SSB Module on ELIC.} 
    \label{fig:ablation-elic}
\end{figure}

\begingroup
\color{black}
\subsection{Algorithm  Independence}
To evaluate the performance of the proposed shift module across different algorithms, we replaced all 3x3 convolutions in the ELIC\cite{he2022elic} model's backbone network with SSB and compared the performance and complexity, as shown in Figure \ref{fig:ablation-elic} and Table \ref{table:ablation-elic}. ELIC-shift is the model after replacement, and its performance loses 2.35\% BD-rate compared to ELIC, but  the computational load is reduced by 133KMACs/Pixel. This indicates that our proposed SSB module is applicable to different networks. Future work could explore applications in other tasks, such as image captioning\cite{MA2023109420}.
\endgroup

\begin{figure}[ht]
    \centering 
    \includegraphics[width=0.40\textwidth]{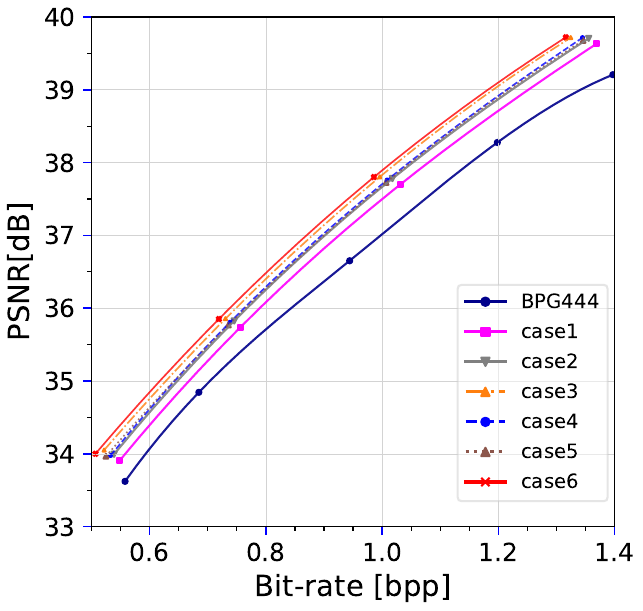}  
    \vspace{-0.5em}
    \caption{\textbf{RD curves from Ablation Studies.} The configuration of these cases are shown in Table \ref{table:ablation_all}. } 
    \label{fig:ablation}
\end{figure}

\section{Ablation Studies}
\label{sec:ablation}
In this section, in order to understand the role of each component, we provide more experimental results and detailed discussions on the two components, spatial shift and channel recursive fusion.
All ablation experiments were performed on the medium model. All the results are obtained using the standard Kodak test dataset.

\subsection{Spatial Shift}
\subsubsection{The Impact of Spatial-Shift}

The impact of shift can be demonstrated by removing the shift operation from the model. As shown in Table \ref{table:ablation_all}, the BD rate deteriorated by 6.23\% after the removal of the shift in the shifting block, indicating the significant contribution of the shift operation. Moreover, the removal of the shift did not reduce the number of parameters, computational load, or the encoder-decoder latency of the model, suggesting that shifting is a cost-effective yet efficient operation. 
Additionally, the $1\times1$ convolution following the shift operation is also crucial. After removing the $1\times1$ convolution, the BD rate worsened by approximately 4.07\%. This may be because the $1\times1$ convolution after shifting effectively enlarges the receptive field of the features, which is beneficial for feature extraction.

\subsubsection{Number of SSB}
In the backbone network, each upsampling/downsampling stage utilizes multiple Spatial-Shift Blocks (SSB) for feature extraction. The more the number of these modules, the greater the performance improvement, but this also results in an increase in the model's computational complexity and number of parameters. Therefore, it's necessary to balance the number of SSBs between performance and complexity.
As shown in Table \ref{table:ablation1}, increasing the number of SSBs reduces the R-D loss but also increases the number of parameters and computational complexity. To keep the complexity of the entire encoder-decoder under 200 KMAC (Kilo Multiply-Accumulate Operations) per pixel, we chose this hyperparameter to be 3.

\begin{table}[t]    
    \caption{{Ablation studies about the SSB and 3x3 Conv.}}
    \label{table:ablation-ssb}
    \centering
    \renewcommand\arraystretch{1.2}
    \setlength{\tabcolsep}{3pt}{ 
    \normalsize
    \begin{tabular}{ccccccc}
    \toprule
    Methods  &R-D Loss$\downarrow$ &Para($M$) &FLOPs  & \makecell{Time\\(GPU)}  &\makecell{Time\\(CPU)}  \\ \hline
    SSB   &2.028   &5.79   &173.34   &147 &737\\
    3x3 Conv   &2.048   &12.92   &618.59  &131 &1640\\
    SSB w/o shift   &2.093  &5.79   &173.34   &137 &393\\
    \bottomrule
    \end{tabular}
    }

\end{table}

\subsubsection{Number of groups of spatial shifts}
Spatial shifting involves a feature grouping operation, which brings diversity to the shifting process. However, feature grouping entails memory access, and too much grouping can disrupt the original distribution of features. Therefore, it is essential to choose an appropriate number of groups to balance performance.
As shown in Table \ref{table:ablation3}, we compared the R-D loss and inference time for different numbers of groups. When the number of groups is set to 4, the best balance between performance and inference time is achieved. This choice provides an optimal trade-off, enhancing the model's efficiency without significantly compromising its effectiveness.\par

\begin{table}[t]
    \caption{Performance of the SSB Module on ELIC.}
    \label{table:ablation-elic}
    \centering
    \renewcommand\arraystretch{1.2}
    \setlength{\tabcolsep}{5pt}{ 
    \normalsize
    \begin{tabular}{ccccc}
    \toprule
    Methods  &BD-rate$\downarrow$ &Para($M$)   &FLOPs(KMACs/Pixel) \\ \hline
    ELIC   &-26.21  &33.79   &833.1   \\
    ELIC-Shift   &-23.86   &32.63  &707.4  \\
    \bottomrule
    \end{tabular}
    }
\end{table}

\begin{table}[t]
    \caption{Ablation studies about the shift step in SSB.}
    \label{table:ablation2}
    \centering
    \renewcommand\arraystretch{1.2}
    \setlength{\tabcolsep}{5pt}{ 
    \normalsize
    \begin{tabular}{ccccc}
    \toprule
    Shift Step  &R-D Loss$\downarrow$ &Para($M$)   &FLOPs(KMACs/Pixel) \\ \hline
    1   &2.0358   &5.79   &173.34   \\
    2   &2.1232   &5.79   &173.34  \\
    3   &2.1349   &5.79   &173.34   \\
    4   &2.1828  &5.79   &173.34  \\
    \bottomrule
    \end{tabular}
    }
\end{table}

\begin{table}[t]
    \caption{Ablation studies about the number of SSB in the backbone network.}
    \label{table:ablation1}
    \centering
    \normalsize
    \renewcommand\arraystretch{1.2}
    \setlength{\tabcolsep}{6pt}{ 
    \begin{tabular}{ccccc}
    \toprule
    Numbers  &R-D Loss$\downarrow$ &Para($M$)   &FLOPs(KMACs/Pixel)\\ \hline
    1   &2.0698   &4.84  & 113.22   \\
    2   &2.0489   &5.31  & 143.28   \\
    3   &2.0358   &5.79  & 173.34   \\
    4   &2.0279   &6.27  & 203.55  \\
    \bottomrule
    \end{tabular}
    }
\end{table}

\subsubsection{Shift Step}
The step size of the spatial shift determines the receptive field of the SSB module; the larger the step size, the larger the receptive field obtained. However, this leads to larger differences in the features before and after processing, and the
Our motivation is to preserve some of the similar features through the shift operation, so larger shifts may be unsuitable for the image compression task. 
As shown in Table \ref{table:ablation2}, the experimental results show that larger step sizes instead lead to lower performance, probably because large step sizes do not guarantee the similarity between features. Thus, we finally chose a step size of 1 for the spatial shift.

\begingroup
\color{black}
\subsection{Shift Operation Time Consumption}
To test the time consumption of the shift operation during the inference process, Table \ref{table:ablation-ssb} shows the performance when using different feature extraction modules in the middle model. The SSB has half the parameters of the 3$\times$3 convolution and approximately one-third of the computational load, and its RD performance is superior to the 3$\times$3 convolution, demonstrating the advantages of SSB. Furthermore, SSB also has advantages in inference time because it only uses 1$\times$1 convolutions, avoiding large kernel convolutions. However, the shift operation introduces additional inference time. Removing the shift reduces inference time on the CPU by 343ms, possibly because the shift operator is not optimized for the CPU.
\endgroup

\begin{table}[t]
    \caption{Ablation studies about the location of CRA.}
    \label{table:ablation5}
    \centering
    \normalsize
    \renewcommand\arraystretch{1.1}
    \setlength{\tabcolsep}{4pt}{ 
    \begin{tabular}{lccc}
    \toprule
    \multicolumn{1}{c}{Location}  &R-D Loss$\downarrow$ &Para($M$)   &FLOPs  \\ \hline
    \begin{tabular}[c]{@{}c@{}}\ding{172}Behind the last SSB\\ and the bottleneck\end{tabular}   &2.0422      &6.44  &162.20   \\ \hline    
    \ding{173}Behind each SSB   &2.0129      &6.08  &246.43   \\   
    \ding{174}Behind the last two SSBs  &2.0358      &5.79  &173.34   \\
    \bottomrule
    \end{tabular}
    }
\end{table}

\begin{table}[t]
    \caption{Ablation studies about groups of spatial shifts.}
    \label{table:ablation3}
    \vspace{-0.5em}
    \centering
    \renewcommand\arraystretch{1.2}
    \setlength{\tabcolsep}{6pt}{ 
    \normalsize
    \begin{tabular}{cccccc}
    \toprule
    Groups  &R-D Loss$\downarrow$ &Para($M$)   &FLOPs (KMACs/Pixel) \\ \hline
    2   &2.0639     &5.79   &173.34   \\
    4   &2.0293     &5.79   &173.34   \\
    8   &2.0358     &5.79   &173.34   \\
    16  &2.0340     &5.79   &173.34   \\
    \bottomrule
    \end{tabular}
    }
    \end{table}

\begin{table}[h]
    \caption{Ablation studies about sampling methods.}
    \label{table:ablationsampling}
    \centering
    \renewcommand\arraystretch{1.1}
    \setlength{\tabcolsep}{4pt}{ 
    \normalsize
    \begin{tabular}{ccccccc}
    \toprule
    Methods  &BD-rate$\downarrow$ &Para($M$)   &FLOPs & \makecell{Time\\(GPU)}  &\makecell{Time\\(CPU)}   \\ \hline
    Shuffle   &2.028   &5.79   &173.34   &130 &774\\
    Conv/TConv   &2.039   &6.44   &244.29  &131 &815\\
    \bottomrule
    \end{tabular}
    }
\end{table}

\subsection{Channel Recursive Attention}

\subsubsection{The Impact of Channel Recursive}
By removing components from the model, the role of channel recursive  attention can be elucidated in detail. As indicated in Table \ref{table:ablation_all}, the BD rate worsened by approximately 1.24\% after removing the channel recursive  fusion operation from the attention mechanism. This deterioration can be attributed to the fact that shift fusion assists in integrating information across the four branches.
Furthermore, eliminating the Spatial Shift Block from the attention mechanism resulted in a decline of about 3.35\% in the BD rate. This may be due to the features entering the shift block undergoing point-wise multiplication operations without mixing spatial information, whereas the shift block aids in the fusion of information in the spatial domain.
We also observed that completely removing the CRA module led to a worsening of the BD rate by approximately 3.51\%. This might be because the attention module facilitates the acquisition of global information, which is beneficial for compression.

\begin{table}[t]
    \caption{Ablation studies about the groups of channel shuffle.}
    \label{table:ablation4}
    \centering
    \renewcommand\arraystretch{1.2}
    \setlength{\tabcolsep}{6pt}{ 
    \normalsize
    \begin{tabular}{cccccc}
    \toprule
    Groups  &R-D Loss$\downarrow$ &Para($M$)   &FLOPs(KMACs/Pixel)  \\ \hline
    2   &2.0358     &5.79   &173.34   \\
    4   &2.0363     &5.79   &173.34   \\
    8   &2.0327     &5.79   &173.34   \\
    16  &2.0377     &5.79   &173.34   \\
    \bottomrule
    \end{tabular}
    }
    \end{table}

\begin{figure}[ht]  
    \centering 
    \includegraphics[width=8.5cm]{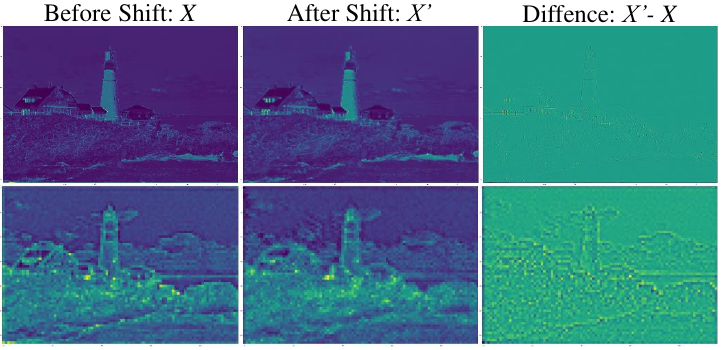}
    \caption{Visualizing spatial shift. Visualize the features before and after the shift operation. The feature data is averaged along the channel dimension. The ``difference" generated before and after the shift corresponds to the edges and contours, consistent with Figure \ref{fig:fig-motivation}. Please zoom in to view.} 
    \label{fig:shift role1}
    \end{figure}

\subsubsection{Location of Attention}
Due to the presence of features of various sizes in the backbone network, the position of attention within the network can affect its computation. The entire network is structured as a bottleneck layer, with higher computational demands near the input and reconstructed image positions.
We compared the complexity and R-D performance of the algorithm when the CRA is placed at three different positions in the backbone network, as shown in Table \ref{table:ablation5}.
When there are 6 CRAs in the backbone network, i.e., one following each SSB, the best performance is achieved. However, at this configuration, the computational load reaches 246 KMACs/Pixel.
With 4 CRAs in the backbone network, positioned at the last two SSBs, the performance is superior to that when positioned at the first location.\par
Considering these findings, we ultimately chose the placement as depicted in Figure \ref{fig:framework}. This decision balances the need for effective attention mechanisms while managing the computational demands of the network.

\subsubsection{The Number of Groups Shuffled}
In channel shuffle, the number of groups determines the extent of feature shuffling. A higher number of groups leads to richer information in each feature group, but it can also disrupt the original feature structure. Thus, determining the optimal number of groups is essential.
Table \ref{table:ablation4} illustrates the performance impact of different group numbers in channel shuffling. When the number of groups is set to 8, the R-D performance is optimal, achieving the best balance. This configuration allows for effective feature shuffling, enhancing the model's ability to capture and represent information efficiently.

\begingroup
\color{black}
\subsection{Sampling}
Table \ref{table:ablationsampling} shows the impact of different sampling methods, where Conv/Tconv represents sampling using 3x3 convolutions and transposed convolutions. From the table, it can be seen that using pixel shuffle for sampling outperforms the convolution method, with advantages in both parameter count and computational load. In terms of inference time, pixel shuffle sampling also has an advantage as it avoids large kernel convolutions. This paper recommends using the pixel shuffle method for sampling in the backbone network.\par
\endgroup

\section{Conclusion}
\label{sec:Conclusion}

In this paper, we propose a lightweight backbone network for LIC, named ShiftLIC. We first examine the redundancy in current LIC algorithms and find a large number of similar features in the backbone network. Based on this, we suggest extensively using shift operations in the backbone network to construct an efficient network.
ShiftLIC achieves an outstanding trade-off in R-D (Rate-Distortion) performance-complexity and RD performance-decoding delay. 
ShiftLIC sets a new SOTA benchmark with a BD-rate gain per MACs/pixel of -102.6\%, showcasing its potential for practical deployment in resource-constrained environments.
Our work reveals the applicability of shift operations to LIC and proposes a lightweight backbone network, which significantly advances the rate-distortion-computation frontier and represents the best current option for performance-complexity trade-offs, with potential for mobile deployment.\par
In the future, we plan to further explore other lightweight strategies, such as quantization and pruning, to aid in the practical deployment of LIC.

\bibliographystyle{IEEEtran}
\bibliography{egbib}

\vfill
\end{document}